\documentclass[12pt]{article}

\usepackage[body={17cm, 23cm, centered}]{geometry}
\usepackage{epsfig,verbatim,cite}
\usepackage{lmodern}
\usepackage{braket}
\usepackage[T1]{fontenc}
\usepackage[utf8]{inputenc}			

\usepackage{mathtext,marvosym,textcomp}
\usepackage{mathtools}
\usepackage{slashed}
\usepackage[usenames,dvipsnames,svgnames,table]{xcolor}
\usepackage{tikz}
\usepackage[ normalem]{ulem}
\usepackage{tocloft}
\usepackage{epsfig,amsmath,amsfonts,amssymb,arydshln}
\usepackage[makeroom]{cancel}
\usepackage{multirow}

 \usepackage[debug,pageanchor=false]{hyperref}
\hypersetup{colorlinks=true,linktocpage,breaklinks,
            urlcolor=blue,
            linkcolor=red,
            citecolor=blue
            }

\numberwithin{equation}{section}

\def\a{\alpha} 
\def\b{\beta} 
\def\g{\gamma} 
\def\d{\delta} 
\def\e{\epsilon}

\def\k{\kappa} 
\def\l{\lambda} 
\def\m{\mu}
\def\n{\nu} 
\def\x{\xi} 
\def\r{\rho}
\def\q{\theta}
\def\s{\sigma} 
\def\t{\tau}  
\def\f{\phi}

\def\w{\omega}

\def\D{\Delta}

\def\L{\Lambda} 
 
\def\X{\Xi} 
 
\def\S{\Sigma}

\def\W{\Omega}

\def\fr{\frac}

\def\dt{\partial}

\def\ph{\phantom}

\def\mc{\mathcal}

\def\mF{\mathcal{F}}

\def\mL{\mathcal{L}}

\def\mQ{\mathcal{Q}}

\def\mP{\mathcal{P}}

\def\tW{\widetilde{W}}

\def\DD{{\mathcal{D}}}

\def\XX{\mathbb{X}}

\def\SS{\mathbb{S}}

\def\o{\bar}

\def\AdS{\mathrm{AdS}}

\def\rmGL{\mathrm{GL}}
\def\rmSO{\mathrm{SO}}
\def\rmSL{\mathrm{SL}}

\def\rmO{\mathrm{O}}

\def\rmE{\mathrm{E}}

\usepackage[pdftex]{pict2e}
\usepackage[dvipsnames]{xcolor}

\usepackage{tcolorbox}

\begin{document}

\renewcommand{\contentsname}{}
\renewcommand{\refname}{\begin{center}References\end{center}}
\renewcommand{\abstractname}{\begin{center}\footnotesize{\bf Abstract}\end{center}} 

\begin{titlepage}
\ph{preprint}

\vfill

\begin{center}
   \baselineskip=16pt
   {\large \bf Tri-vector deformations with external fluxes  
   }
   \vskip 2cm
    Sergei Barakin,$^\dagger{}^\star$\footnote{\tt barakin.serge@gmail.com}
    Kirill Gubarev,$^\dagger$\footnote{\tt kirill.gubarev@phystech.edu}
    Edvard T. Musaev$^{\dagger \star \bullet }$\footnote{\tt musaev.et@phystech.edu}
       \vskip .6cm
             \begin{small}
                          {\it
                          $^\dagger$Moscow Institute of Physics and Technology, 
                          Laboratory of High Energy Physics\\
                          Institutskii pereulok, 9, 141702, Dolgoprudny, Russia\\[0.5cm]
                          $^\star$Institute of Theoretical and Mathematical Physics, Lomonosov Moscow State University, Moscow 119991, Russia                      
                          } \\ 
\end{small}
\end{center}

\vfill 
\begin{center} 
\textbf{Abstract}
\end{center} 
\begin{quote}

We extend the formalism of tri-vector deformations to the full SL(5) exceptional field theory with no truncation assumed thus covering 11D backgrounds of any form. We derive explicit transformation rules for 11D supergravity component fields and prove that these generate solutions given the same algebraic conditions hold: generalized Yang--Baxter equation and the unimodularity condition.
\end{quote} 

\vfill
\setcounter{footnote}{0}
\end{titlepage}

\tableofcontents

\setcounter{page}{2}

\section{Introduction}

Among various approaches to quantum field theory gauge/gravity duality seems to be the most promising to gain access to non-perturbative effects. However, in its strongest form AdS/CFT correspondence is restricted to highly symmetric theories such as the $\mc{N}=4$ $d=4$ super Yang-Mills theory dual to string theory on the AdS${}_5\times \SS^5$  background \cite{Maldacena:1997re}. Certainly of most interest are theories that exhibit less supersymmetry, non-conformal, at finite temperature etc, which for obvious reasons evade methods and approaches based on (super)symmetric arguments. It is however possible to keep the same control of a subset of such theories described as various deformations of highly (super)symmetric conformal field theories: by adding (ir)relevant or marginal operators, changing the microscopic behaviour by introducing non-commutativity, or by considering a 6phase with a non-vanishing VEV of an operator. Among the most well known examples one finds Leigh-Strassler $\mc{N}=1$ $d=4$ SCFT that is a deformation of the $\mc{N}=4$ $d=4$ SYM by marginal and relevant operators \cite{Leigh:1995ep}, and Argyres-Douglas theories describing a non-Lagrangian Coulomb phase of an $\mc{N}=2$ SCFT's with operators of fractional dimensions \cite{Argyres:1995jj}. While the latter have also been recently assigned a holographic dual \cite{Bah:2021mzw}, directly relevant to this paper is the holographic description of exactly marginal Leigh-Strassler deformations. These correspond to families of supergravity solutions of the form AdS${}_5\times M_5$ where $M_5$ is a 5-dimensional Riemannian manifold, that is topologically a 5-sphere, obtained by a bi-vector Yang-Baxter deformation of the $\AdS_5\times \SS^5$ solution. Such deformations have been long known in the context of 2d sigma-models for preservation of integrability \cite{Delduc:2013qra,Arutyunov:2015qva} and uplifted to the full supergravity framework in \cite{Bakhmatov:2017joy,Bakhmatov:2018apn} thus relating to the earlier work \cite{Lunin:2005jy} of Lunin and Maldacena where the holographic dual of the Leigh-Strassler $\beta$-deformation was found.

In this framework the bi-vector is constructed of Killing vectors, that in the case of AdS backgrounds can be taken i) all along the AdS space, ii) all along the internal space (say, the sphere) or iii) mixed. In the first case on ends up with a non-commutative dual theory (see e.g. \cite{vanTongeren:2015uha}), in the second case the deformation is dual to adding of exactly marginal operators, while the latter case describes dipole deformations (see e.g. \cite{Guica:2017mtd}). The brane picture has been developed in \cite{Imeroni:2008cr}. Since the internal space of relevant for holography backgrounds is compact, the ability of bi-vector Yang-Baxter transformations to generate exactly-marginal deformation on the field theory side is highly restricted by the theorem \cite{Lichnerowicz:1988abc,Pop:2007abc}. The theorem states that classical Yang-Baxter equation for algebra of a compact group (real) has non-trivial solutions only when the group is abelian. As it has been shown in \cite{Musaev:2023own} by presenting an explicit example such a theorem does not exist for the generalization of the classical Yang-Baxter equation (genCYBE) found in \cite{Sakatani:2019zrs,Malek:2019xrf}, that regulates tri- and six-vector transformations of 11-dimensional backgrounds. Early examples of tri-vector deformations have been constructed in \cite{Ashmore:2018npi}, using special geometry, and in \cite{Lunin:2005jy},  using SL(2,$\,\mathbb{R}$) group for abelian isometries. The general approach to polyvector deformations, that has been suggested in \cite{Bakhmatov:2019dow} and further developed in the series of papers \cite{Bakhmatov:2020kul,Gubarev:2020ydf,Gubarev:2024tks}, is based on exceptional field theory where supergravity fields are collected into an irrep of the corresponding U-duality group. 

The formalism of polyvector deformations developed in \cite{Gubarev:2020ydf} is restricted to backgrounds of the form $M_D \times N_{11-D}$. To be more specific: the fields are explicitly split into those depending on coordinates $x^\m$ on the $D$-dimensional manifold $M_D$ and those depending on the rest $11-d$ coordinates $x^m$; the 11d metric does not have mixed $\m m$ components; all vector and tensor fields on $M_D$ vanish. Such a truncation, although sufficiently simplifies technicalities behind the calculations, severely restricts the set of deformations covered by the framework. For example, it is possible to deform say the $\AdS_7\times \SS^4$ background along the isometries of the 4-sphere using the SL(5) formalism, however one cannot use it for isometries of the AdS part (certainly, one could construct a $E_7$ generalization, that would cover the 7-dimensional AdS space). In this paper, we aim at further development of the formalism of polyvector deformations such that it covers 11d backgrounds in the 7+4-split with tensor fields. We present explicit transformation rules for 11d fields and condition sufficient for them to generate a solution. The algebraic condition is again the generalized Yang-Baxter equation of \cite{Sakatani:2019zrs,Malek:2019xrf} as expected.

The paper is structured as follows. In Section \ref{sec:sl5} we briefly overview the formalism of the SL(5) exceptional field theory with a particular emphasis at the relation between the SL(5) covariant fields and the fields of the 11D supergravity. In this section we also provide full SL(5) covariant equations of motion in terms of generalized fluxes (in the sense of \cite{Gubarev:2023kvq}) and covariant field strengths. In Section \ref{sec:defs} we define 3-vector deformations as SL(5) transformation generated by Killing vectors and derive sufficient conditions for these to be solution generating. We also derive explicit transformation rules for components of the 11D metric and 3-form field. Section \ref{sec:concl} is devoted to discussion of the results and the remaining open questions.

\section{U-duality covariant form of 11D supergravity}

Poly-vector transformations, including the bi-vector case, are always given by an element of the corresponding duality group. The crucial difference is that bi-vector transformations of a $d$-dimensional subspace belong to the orthogonal (T-duality) group $\rmO(d,d)$ and in principle one can take $d=10$ and consider the most general transformation of a full 10d supergravity solution. Important, that all the fields of the Type II 10d supergravity can be collected into multiplets transforming linearly under bi-vector transformations. 

This is no longer true for poly-vector transformations of 11d backgrounds, these belong to U-duality groups $\rmSL(5), \, \rmSO(5,5), \rmE_6$ etc. depending on the dimension of the subspace whose isometries generate the transformation. Hence, only tensors on this subspace are collected into irreps of the duality group using the procedure described in \cite{Cremmer:1997ct} and transform linearly under the deformation. This implies, that to describe polyvector transformation one necessarily has to consider a split $11=D+d$ and, more importantly, to work with fields collected in multiplets under the duality groups to realize deformations linearly.

Such a formalism is provided by the so-called exceptional field theory (ExFT) that is a U-duality-covariant formulation of supergravity. In this section we provide explicit expressions for the Kaluza-Klein decomposition of 11d supergravity required to construct its embedding into ExFT, and write equations of motion in the U-covariant form to check their invariance un tri-vector deformation subject to certain conditions (genCYBE and the so-called unimodularity condition).

\subsection{Decomposition 7+4 of 11D supergravity}

Let us start with the standard Kaluza-Klein decomposition of the 11D supergravity and all its symmetries, however, keeping dependence on all 11 coordinates. Hence, here we aim at a formulation of the full 11D supergravity manifestly covariant under the $\rmGL(7)\times \rmGL(4)$ diffeomorphisms. Certainly, invariance under the full 11D diffeomorphism group remains and is hidden. Below we mainly follow \cite{Hohm:2013vpa} rewriting expressions for the 7+4 case relevant here, and aim at explicit embedding of the 11D fields into the SL(5) ExFT. The action of the 11D supergravity reads 
\begin{equation} 
\begin{aligned}
    S = \int d^{11}x \left( E \hat{R} - \frac{2}{4!} E \hat{F}^{\hat{\mu}_{1} \dots \hat{\mu}_{4}} \hat{F}_{\hat{\mu}_{1} \dots \hat{\mu}_{4}} + \frac{2}{3^2 \, 4!^2} \epsilon^{\hat{\mu}_{1} \dots \hat{\mu}_{11}} \hat{F}_{\hat{\mu}_{1} \dots \hat{\mu}_{4}} \hat{F}_{\hat{\mu}_{5} \dots \hat{\mu}_{8}} C_{\hat{\mu}_{9} \dots \hat{\mu}_{11}} \right),
    \label{b11SUGRA}
\end{aligned} 
\end{equation}
where $E$ denotes determinant of the full 11D vielbein $E_{\bar \m}{}^{\hat{a}}$, the Ricci scalar $R$ is given in terms of the anholonomy coefficients $\hat{\Omega}_{\hat{a} \hat{b} \hat{c}} $ as follows
\begin{equation} 
\begin{aligned}
    \hat{R} & = E^{\hat{\mu}}{}_{\hat{a}} E^{\hat{\nu}}{}_{\hat{b}} \hat{R}_{\hat{\mu} \hat{\nu}}{}^{\hat{a} \hat{b}} = \frac{1}{4} \hat{\Omega}^{\hat{a} \hat{b} \hat{c}} \hat{\Omega}_{\hat{a} \hat{b} \hat{c}} + \frac{1}{2} \hat{\Omega}^{\hat{a} \hat{b} \hat{c}} \hat{\Omega}_{\hat{b} \hat{c} \hat{a}} + \hat{\Omega}_{\hat{c} \hat{a}}{}^{\hat{a}} \hat{\Omega}^{\hat{c}}{}_{\hat{b}}{}^{\hat{b}}, \\
    \hat{\Omega}_{\hat{a} \hat{b} \hat{c}} &=  E^{\hat{\mu}}{}_{\hat{a}} E^{\hat{\nu}}{}_{\hat{b}} ( \partial_{\hat{\mu}}{E_{\hat{\nu} \hat{c}}} - \partial_{\hat{\nu}}{E_{\hat{\mu} \hat{c}}} ),
\end{aligned} 
\end{equation}
and the 4-form field strength reads $\hat{F}_{\hat{\mu} \hat{\nu} \hat{\rho} \hat{\sigma}} = 4 \partial_{[\mu} C_{\hat{\nu} \hat{\rho} \hat{\sigma}]} $. Our index notations are collected in Appendix \ref{app:index}

The general Kaluza-Klein ansatz for the 7 + 4 decomposition of the vielbein reads
\begin{equation} 
\begin{aligned}
    E_{\hat{\mu}}{}^{\hat{a}} = 
        \begin{pmatrix}
        \phi^{-\fr15} e_\mu{}^{\o{\mu}} & A_\mu{}^m \phi_m{}^{\o{m}} \\
    0                         & \phi_m{}^{\o{m}}                 \end{pmatrix} , &&
    E^{\hat{\mu}}{}_{\hat{a}} = 
        \begin{pmatrix}
        \phi^{\fr15} e^\mu{}_{\o{\mu}} & -\phi^{\fr15} A_\mu{}^m e^\mu{}_{\o{\mu}} \\
        0                          & \phi^m{}_{\o{m}} \end{pmatrix},
\end{aligned} 
\end{equation}
where $\phi = \det (\phi_{m}{}^{\o{m}})$ to end up in the Einstein frame for the 7-dimensional vielbein $e_\m{}^{\o{\m}}$. Certainly the upper-triangular form of the vielbein does not maintain under a general diffeomorphism transformation and an additional rotation $\rmSO(4)$ is required. To be more specific let us first write decomposition of the 11D diffeomorphism parameter $\hat{\x}^{\hat{\m}}$ and of a $\rmSO(1,10)$  Lorentz transformation matrix 
\begin{equation} 
\begin{aligned}
    \hat{\xi}^{\hat{\mu}} = \begin{pmatrix}
        \xi^\mu\\
        \Lambda^m - \xi^\nu A_\nu{}^m
    \end{pmatrix} , &&
    \lambda^{\hat{a}}{}_{\hat{b}} = \begin{pmatrix}
        \lambda^{\o{\mu}}{}_{\o{\nu}} & \lambda^{\o{\mu}}{}_{\o{n}}\\
        \lambda^{\o{m}}{}_{\o{\nu}} & \lambda^{\o{m}}{}_{\o{n}}
    \end{pmatrix} \in \rmSO(1,10).
\end{aligned} 
\end{equation} 
To restore the upper-triangular form of the vielbein after a diffeomorphism one must complement it by $\lambda^{\hat{a}}{}_{\hat{b}}$ with
\begin{equation} 
    \begin{aligned}
    \lambda^{\o{\mu}}{}_{\o{m}} = - \phi^{-\fr15} e_{\mu}{}^{\o{\mu}} \dt_{\o{m}}{\xi^{\mu}}, \quad \lambda_{\o{\mu}}{}^{\o{m}} = \fr15 \phi^{-\fr15} e_{\nu \o{\mu}} \dt^{\o{m}}{\xi^{\nu}}.
\end{aligned} 
\end{equation}
Given that and starting from the full 11D diffeomorphism and Lorentz transformations
\begin{equation}
    \d_{\hat{\x}, \hat{\l}} E_{\hat{\m}}{}^{\hat{a}} = \x^{\hat{\n}} \dt_{\hat{\n}} E_{\hat{\m}}{}^{\hat{a}} + E_{\hat{\n}}{}^{\hat{a}} \dt_{\hat{\m}} \x^{\hat{\n}} + \l^{\hat{a}}{}_{\hat{b}} E_{\hat{\m}}{}^{\hat{b}} ,
\end{equation}
one ends up with the following transformation rules for the fields $e_\m{}^{\o{\m}}$, $A_\m{}^m$ and $\f_m{}^{\o{m}}$:
\begin{equation} 
\begin{aligned}
    \d e_\m{}^{\o{\m}}  & = \xi^\n D_\n e_\m{}^{\o{\m}} + e_\n{}^{\o{\mu}} D_\m \xi^\n + \mL_\L e_\n{}^{\o{\n}} + \l^{\o{\m}}{}_{\o{\n}} e_\m{}^{\o{\nu}}  \\
    \d A_\mu{}^m & = \xi^\n F_{\nu \mu}{}^m + D_\m \L^m \\
    \d \f_m{}^{\o{m}} &= \xi^\n D_\n \phi_m{}^{\o{m}} + \mL_\L \phi_m{}^{\o{n}} + \l^{\o{m}}{}_{\o{n}} \phi_m{}^{\o{n}}.
\end{aligned} 
\end{equation}
These are written in the form of covariant transformations of 7D tensor fields and the long derivative $D_\m$ and the field strength $F_{\m\n}{}^m$ are defined as  follows
\begin{equation} 
\begin{aligned}
    D_\m e_\n{}^{\o{\n}} & = \left( \dt_\m - A_\m{}^n \dt_n \right) e_\n{}^{\o{\n}} - \fr15 (\dt_n A_\m{}^n) e_\n{}^{\o{\n}} \\
    D_\mu \phi_m{}^{\o{n}} & = \left( \dt_\m - A_\m{}^n \dt_n \right) \phi_m{}^{\o{n}} - \phi_n{}^{\o{n}} \dt_m A_\m{}^n \\
    F_{\mu \nu}{}^{m} & = 2 \partial_{[\m} A_{\n]}{}^m - A_{[\m}{}^n \dt_n A_{\n]}{}^m .
\end{aligned}
\end{equation}

Similarly, can be written the Kaluza-Klein decomposition for the 3-form field $C_{\hat{\m}\hat{\n}\hat{\rho}}$. It is convenient to define tensor fields $A_{\dots}$ as one, covariant transforming under 11D diffeomorphisms:
\begin{equation} 
\begin{aligned}
    A_{\mu \nu \rho} &= C_{\mu \nu \rho} - 3 A_{[\mu}{}^{m} C_{\nu \rho] m} + 3 A_{[\mu}{}^{m} A_{\nu}{}^{n} C_{\rho] m n} - A_{\mu}{}^{m} A_{\nu}{}^{n} A_{\rho}{}^{k} C_{m n k}\\
    A_{\mu \nu m} & = C_{\mu \nu m} - 2 A_{[\mu}{}^{n} C_{\nu] m n} + A_{\mu}{}^{n} A_{\nu}{}^{k} C_{m n k}  \\
    A_{\mu m n} & = C_{\mu m n} - A_{\mu}{}^{k} C_{k m n} \\
    A_{m n k} & = C_{m n k}.
\end{aligned} 
\end{equation}
Components of the 11-dimensional 4-form field strength are written in terms of the covariant potentials and the derivative  $D_{\mu} = \partial_{\mu} - \mc{L}_{A_\mu{}^m}$ as follows 
\begin{equation} 
\begin{aligned}
    F_{m n k l} & = 4\partial_{[m}{A_{n k l]}}, \\
    F_{\mu m n k} & = D_{\mu}{A_{m n k}} - 3 \partial_{[m}{A_{|\mu| n k]}}, \\
    F_{\mu \nu m n} &= 2 D_{[\mu}{A_{\nu] m n}} + F_{\mu \nu}{}^{k} A_{k m n} + 2 \partial_{[m}{A_{|\mu \nu| n]}}, \\
    F_{\mu \nu \rho m} &= 3 D_{[\mu}{A_{\nu \rho] m}} + 3 F_{[\mu \nu}{}^{n} A_{\rho] m n} - \partial_{m}{A_{\mu \nu \rho}}, \\
    F_{\mu \nu \rho \sigma} &= 4 D_{[\mu}{A_{\nu \rho \sigma]}} + 6 F_{[\mu \nu}{}^{m} A_{\rho \sigma] m} .
\end{aligned} 
\end{equation}

In terms of these covariant fields the Einstein-Hilbert and tensor kinetic parts of the full 11-dimensional supergravity Lagrangian are written as follows
\begin{equation} 
\label{eq:EHkin_decomposed}
\begin{aligned}
    e^{-1} E \hat{R}  = & \,   R[g] - \frac{1}{4} \phi^{\fr25} F^{\mu \nu}{}_{m} F_{\mu \nu}{}^{m} + V_{E H}(\phi , e) \\ 
    & - \frac{1}{2} g^{\mu \nu} \left( \phi^{m n} {D}_{\mu}{\phi_{m}^{\alpha}} {D}_{\nu}{\phi_{n \alpha}} + ( \phi^{\alpha m} {D}_{\mu}{\phi_{m}{}^{\beta}} ) ( \phi_{\beta}{}^{n} {D}_{\nu}{\phi_{n \alpha}} ) +\fr25 \phi^{-2} {D}_{\mu}{\phi} {D}_{\nu}{\phi} \right) \\
    e^{-1} E & \hat{F}^{\hat{\mu}_1 \dots \hat{\mu}_4} \hat{F}_{\hat{\mu}_1 \dots \hat{\mu}_4} = \phi^{\fr65} F^{\mu \nu \rho \sigma} F_{\mu \nu \rho \sigma} + 4 \phi^{\fr45} F^{\mu \nu \rho m} F_{\mu \nu \rho m} \\
    & + 6 \phi^{\fr45} F^{\mu \nu}{}_{m k} F_{\mu \nu}{}^{m k} + 4 F^{\mu m n k} F_{\mu m n k} + \phi^{-\fr25} F^{m n k l} F_{m n k l}
\end{aligned} 
\end{equation}
where $V_{E H}(\phi , e)$ is the so-called scalar potential and contains only derivatives of 7D scalar fields along 4D coordinates $x^m$
\begin{equation}
\begin{aligned}
    V_{E H}(\phi , e) & = \phi^{-\fr25} \left( R(\phi) + \frac{1}{4} \phi^{m n} ( D_m g^{\mu \nu} D_n g_{\mu \nu} + (g^{-1} D_m g) (g^{-1} D_n g) ) \right) , \\
    D_{m}{e_{\nu}{}^{\o{\nu}}} & = \left( \dt_m - \fr15 \phi^{-1} \dt_m \phi \right) e_{\nu}{}^{\o{\nu}} .
\end{aligned} 
\end{equation}

Upon the toroidal Kaluza-Klein reduction, i.e. $\dt_m=0$, the scalar potential is zero, while for other reduction this gives the scalar potential of the corresponding gauged supergravity, hence the name. The main difference between Chern-Simons terms and the other parts of the Lagrangian is that the former cannot be written in a full SL(5)-covariant form. This is a common feature of exceptional field theories and maximal (un)gauged supergravities, where only variation of Chern-Simons terms are covariant under the U-duality group. Since we are eventually interested in equations of motion, i.e. variations of the action, we need only variation of the Chern-Simons part rather the expression itself. For this reason we will not mention these terms until Section \ref{sec:equations} where we will simply use the expression from \cite{Musaev:2015ces}.

\begin{table}[h!]
  \centering
    \begin{tabular}{ | c | c | c | c | c | c | c | c | c | c | c | }
    \hline
    $d$ &  3 & 4 & 5 & 6 & 7 & 8 & 9 \\
    \hline
    $E_{d(d)}$ & SL(3) $\times$ SL(2) & SL(5) & SO(5,5) & $E_{6(6)}$ & $E_{7(7)}$& $E_{8(8)}$ & $E_{9(9)}$ \\
    \hline
    \end{tabular}
    \caption{ $U$-duality symmetry groups of D+d decomposed 11D SUGRA }
    \label{tab:11DExcGr}
\end{table}

Now, to see a relation between fields of an exceptional field theory transforming linearly under polyvector deformations and the fields of 11D supergravity in the split form, we must specify U-duality group (see Table. \ref{tab:11DExcGr}). Of interest for us here is the case $d = 4$ corresponding to the SL(5) exceptional field theory. In this case, the field content of the split form in terms of 7D tensors reads: one external metric field $g_{\m\n}$, 10 vector fields $A_\m{}^m$ and $A_{\m m n}$, 14 scalar fields $\f_{mn}$ and $A_{mnk}$, four 2-forms $A_{\m\n m}$ and one 3-form $A_{\m\n\r}$. The metric $g_{\m\n}$ is a singlet under the $\rmSL(5)$ group, the scalars combine into 
\begin{equation}
\begin{aligned}
     \mc{E}_{M}{}^{\bar{M}} & = \phi^{\frac{3}{5}} \begin{pmatrix}
        \phi^{-1} \phi_{m}{}^{\bar{m}} & - V^{n} \phi_{n}{}^{\bar{m}} \\
        0 & 1
    \end{pmatrix}
    \in \fr{\rmSL(5)}{\rmSO(5)} ,
\end{aligned}
\end{equation}
where $ V^{m} = 1/3! \, \phi^{-1} \, e^{mnkl}A_{nkl} $ and the indices $M,\bar{M} =1,\dots,5$ label the $\mathbf{5}$ of SL(5). Note that $\e^{mnkl}$ is the epsilon-symbol (does not include $\f$). This will be called the generalized vielbein of exceptional field theory, and the distinction between barred and unbarred capital Latin indices is the same as between barred and unbarred small Latin indices (flat and curved). The corresponding generalized metric is given by 
\begin{equation}
    m_{MN} = \mc{E}_M{}^{\bar{M}}\mc{E}_N{}^{\bar{N}}m_{\bar{M} \bar{N}},
\end{equation}
where $m_{\bar{M} \bar{N}}$ is the flat (diagonal) metric. Note that $\mc{E}_M{}^{\bar{M}}$ and $m_{MN}$ are the standard generalized vielbein and generalized metric of the SL(5) ExFT. To turn to flux formulation and to define generalized fluxes invariant under polyvector deformations one has to rescale the metric as \cite{Gubarev:2020ydf}  
\begin{equation}
    E_M{}^{\bar{M}} = e_{(7)}^{-\fr{1}{14}}\mc{E}_N{}^{\bar{M}},
\end{equation}
where $e_{(7)}=\det e_\m{}^{\bar{\m}}$ is determinant of the external ExFT metric $g_{\m\n}$.

Vector fields combine into the 10-dimensional irrep of $\rmSL(5)$ as follows
\begin{equation}
    A_\mu{}^{[M N]} =
        \begin{pmatrix}
            A_\mu{}^m \\
            \fr12 \, \e^{mnkl}A_{\mu k l}.
        \end{pmatrix} 
\end{equation}
For the 2- and 3-forms the story is a bit more complicated since we have four of the former and one of the latter which on themselves do not combine into irreps of SL(5). Instead, following \cite{Cremmer:1997ct} one has to dualise tensor fields to the lowest possible rank. To do that in consistency with notations of the SL(5) ExFT let us first introduce extra fields $\tilde{A}_{\m\n}$ and $\tilde{A}_{\m\n\r}{}^m$ and define
\begin{equation}
    B_{\m \n M} = 
        \begin{pmatrix}
            - 8 ( A_{\m \n m} + A_{[\mu}{}^n A_{\nu] m n} ) \\
            \tilde{A}_{\m\n}
        \end{pmatrix}, \quad
    C_{\m \n \r}{}^M = 
        \begin{pmatrix}
            \tilde{A}_{\m\n\r}{}^m \\
            8 ( A_{\m\n\r} + 3 A_{[\m}{}^m A_\n{}^n A_{\r] m n} )
        \end{pmatrix}.
\end{equation}
The fields $\tilde{A}_{\m\n}$ and $\tilde{A}_{\m\n\r}{}^m$ are new fields introduced for the sake of covariance and are related to the Lagrange multipliers introduced in \cite{Cremmer:1997ct} to render covariant equations. For the fields defined above, the field strengths composed as follows
\begin{equation}
    \begin{aligned}
    \mathcal{F}_{\mu \nu}{}^{[M N]} & =
    \begin{cases}
        \mF_{\mu \nu}{}^{5 m} = F_{\mu \nu}{}^m, \\
        \mF_{\mu \nu}{}^{m n} = \fr12 \, \e^{m n k l} \left( F_{\mu \nu k l} - F_{\mu \nu}{}^{p} A_{k l p} \right),
    \end{cases}\\
    \mF_{\m\n\r M} & =
    \begin{cases}
        \mF_{\m\n\r 5} = \phi \left( \fr{e}{4!} \epsilon_{\mu \nu \rho \sigma \eta \zeta \theta} \mF^{\sigma \eta \zeta \theta 5} \phi^{\fr15} - V^m \mF_{\mu \nu \rho m} \right) \\
        \mF_{\mu \nu \rho m} = - 8 ( F_{\mu \nu \rho m} + F_{[\mu \nu}{}^n A_{\rho] n m} )
    \end{cases}  \\
    \mF_{\m\n\r\s}{}^M & =
    \begin{cases}
        \mF_{\mu \nu \rho \sigma}{}^{5} = 8 F_{\mu \nu \rho \sigma} , \\
        \mF_{\m\n\r\s}{}^m = \phi \left( \mF_{\m\n\r\s}{}^5 V^{m} + \fr{e}{3!} \epsilon_{\mu \nu \rho \sigma \eta \zeta \theta} \phi^{m n} \mF^{\eta \zeta \theta}{}_{n} \phi^{-\fr15} \right)
    \end{cases}  
    \end{aligned}
\end{equation}
match those coming from the SL(5) exceptional field theory (see further in \eqref{eq:hierarchy}) and are required to satisfy the following self-duality condition
\begin{equation}
    e \, m^{MN} \mF^{\m\n\r}{}_{N} = \frac{1}{4!} \e^{\m\n\r\l\s\t\k} \mF_{\l\s\t\k}{}^M,
\end{equation}
where $e = (-\det g_{\m\n})^{1/2}$. These five equations relate the extra five fields to $A_{\m \n m} $ and $A_{\m\n\r}$ up to gauge transformations. Eventually we end up with the following SL(5)-covariant field content of the $11=7+4$ split supergravity
\begin{equation}
    \label{eq:fieldcontent}
    \begin{aligned}
         & g_{\m\n}, && A_\m{}^{MN}, && m_{MN}, && B_{\m \n M}, && C_{\m\n\r }{}^{M}.
    \end{aligned}
\end{equation}
Here $m_{MN} = E_M{}^{\bar{M}}E_N{}^{\bar{N}}m_{\bar{M}\bar{N}}$ and $m_{\bar{M}\bar{N}}$ is the unit matrix. These fields will transform linearly under tri-vector deformations, and to check invariance of equations of motion of 11D supergravity one has to write these in an SL(5)-covariant form. The most straightforward way to do that is to start with equations of the SL(5) exceptional field theory, perform symmetry reduction $\rmSL(5) \to \rmGL(4)$ and use the above field identifications. Before switching to the discussion of exceptional field theory, it is worth to mention that the discussion of \cite{Bakhmatov:2020kul,Gubarev:2020ydf} has been limited to the case when only $g_{\m\n}$ and $m_{MN}$ non-vanishing. In \cite{Gubarev:2020ydf} there we introduced special scalar fluxes, through which it is possible to complete the observed ExFT SL(5) section:
\begin{equation} 
\begin{aligned}
    \mc{F}_{\mu \o{M}}{}^{\o{N}} = & E^M{}_{\o{M}} \mc{D}_{\mu} E_{M}{}^{\o{N}} = \begin{cases}
        \mF_{\m \o{m}}{}^{\o{n}} = - \phi^{-\fr25} \phi_{k}{}^{\o{n}} \mc{D}_{\mu} \left( \phi^{\fr25} \phi^{k}{}_{\o{m}} \right) , \\
        \mF_{\m \o{5}}{}^{\o{n}} = - \fr16 \phi^{-1} \phi_{m}{}^{\o{n}} \epsilon^{m n k l} F_{\mu n k l} , \\
        \mF_{\m \o{5}}{}^{\o{5}} = \phi^{-\fr35} \mc{D}_{\mu} \phi^{\fr35},
    \end{cases}\\
    Y_{\o{M}  \o{N}} = & - E^M{}_{(\o{M}} \dt_{M N} E^N{}_{\o{N})} = \begin{cases}
        Y_{\o{5} \o{m}} = - \phi^{-\fr65} \partial_{n}\left(\phi \, \phi^{n}{}_{\o{m}}\right) ,\\
        Y_{\o{5} \o{5}} = - \fr1{24} \phi^{-\fr65} \epsilon^{m n k l} F_{m n k l},
    \end{cases}\\ 
    \end{aligned}
\end{equation}

\subsection{Exceptional Field Theory SL(5)}

Exceptional field theory (ExFT) is a formalism for maximally supersymmetric supergravity covariant under a U-duality group from Table \ref{tab:11DExcGr}. To ensure covariance it is not enough to collect fields as in \eqref{eq:fieldcontent} since derivatives $\dt_m$ along coordinates of the $d$-dimensional space break the duality group to its geometric $\rmGL(d)$ subgroup. To overcome that additional coordinates are added to the formalism, usually dubbed dual, winding or non-geometric, completing d-coordinates $x^m$ to a set $\XX^{\mc{M}}$. Here $\mc{M}=1,\dots,\mbox{dim}\mc{R}_V$ and $\mc{R}_V$ is the same irrep under which vector fields of the theory transform. The idea of extending coordinates of the target space-time originates from the equivalence of momentum and winding modes of a closed string on a torus, that stands behind T-duality \cite{Buscher:1987qj,Buscher:1987sk}. The corresponding doubling of the space-time has been first considered in \cite{Fradkin:1984ai} in terms of a CFT, and further in \cite{Siegel:1993th,Siegel:1993xq} in terms of local coordinate symmetries (generalized Lie derivatives). The O(10,10)-covariant action of 10-dimensional Type II supergravity was formulated in \cite{Hohm:2010pp} in terms of generalized metric. Although the interpretation in terms of winding modes for a membrane is rather elusive a similar framework can be constructed for U-duality groups, however, a split of 11 dimensions is required. For the scalar sector with SL(5) symmetry this has been done in \cite{Berman:2010is}, the full SL(5) theory that includes tensor fields has been constructed in \cite{Musaev:2015ces} following the general idea formulated in \cite{Hohm:2013pua,Hohm:2013vpa}. 

Let us now stick to the SL(5) case where $D=7$, $d=4$ and the extended space is parametrised by coordinates $\XX^{MN} = - \XX^{NM}$. Since the framework of exceptional field theory is only needed here to conveniently write equations of the 11D supergravity in a covariant form we will always assume $\dt_{mn} = 0$, i.e. the only non-vanishing derivatives are those along the standard geometric coordinates $\dt_{5m} = \dt_m$. However, to keep explicit covariance of all expressions we will be using $\dt_{MN}$. Symmetries of the theory are defined by the so-called generalized Lie derivatives, that for a vector $V^M$ of arbitrary weight $\l[V^M]$ are given by 
\begin{equation}
\label{vectGL}
    \mL_{\L}V^{M} = \frac{1}{2} \L^{K L} {\dt}_{K L}{{V}^{M}}\,  - V^{L} {\dt}_{L K}{{\L}^{M K}}\, + \bigg(\frac{1}{5} + \frac{\lambda[V^{M}]}{4} \bigg) V^{M} {\partial}_{K L}{{\L}^{K L}}.
\end{equation}
For a covector one has
\begin{equation}
\label{covectGL}
    \mL_{\L} V_{M} = \frac{1}{2} \L^{K L} {\dt}_{K L}{{V}_{M}}\,  + V_{L} {\dt}_{M K}{{\L}^{L K}}\, + \bigg( - \frac{1}{5} + \frac{\lambda[V_{M}]}{4} \bigg) V_{M} {\partial}_{K L}{{\L}^{K L}}.
\end{equation}
Consistency of the derivatives requires the so-called section constraint  $\e^{MNKLP}\dt_{MN}\otimes \dt_{KL} = 0$ that is always satisfied given $\dt_{mn} = 0$. Derivatives $\dt_\mu$ along external coordinates enter via a covariant derivative
\begin{equation}
    \mc{D}_\m = \dt_\m - \mL_{A_\m}.
\end{equation}
Upon the solution $\dt_{mn}=0$ of the section constraint $\mc{D}_\m = D_\m$, which is the reason for all the fields in the split supergravity formulation to be collected into combinations covariant w.r.t. $D_\m$. A commutator of $\mc{D}_\m$'s generates a field strength which appears to be not a tensor under generalized Lie derivative and must be corrected by terms that include the 2-forms $B_{\m\n M}$. The 3-form field strength resulting from the corresponding Bianchi identities also fails to be covariant and has to be corrected by the 3-form $C_{\m\n\r}{}^M$ and so on. This tower is called tensor hierarchy and in principle goes on forever, but can be truncated after the 3-form since its field strength already does not contribute field equations. For a detailed discussion of tensor hierarchy for the SL(5) case see \cite{Gubarev:2023kvq}, general discussion in the context of gauged supergravities can be found e.g. in \cite{deWit:2002vt}. Skipping all that we simply provide the resulting expressions for field strengths
\begin{equation}
\label{eq:hierarchy}
\begin{aligned}
\mF_{\m\n}{}^{MN} =&\ 2 \dt_{[\m}A_{\n]}{}^{MN} - [A_{\m},A_{\n}]_E^{MN} - \frac1{16} \epsilon^{M N K L P} \dt_{K L} B_{\m \n P},\\
\mF_{\m \n \r M}  =&\ 3 \, \DD_{[\m} B_{\n \r] M} + 6 \, \epsilon_{M P Q R S} (A_{[\mu}{}^{P Q} \dt_{\nu} A_{\rho]}{}^{R S} - \frac13 [A_{[\mu},A_{\nu}]_{E}^{P Q} A_{\rho]}{}^{R S}) - \dt_{M N} C_{\m \n \r}{}^{N}, \\
\mF_{\m\n\r\s}{}^{M} =&\ 
4\,\mc{D}_{[\m}C_{\n\r\s]}{}^{M} - 6 \mF_{[\m\n}{}^{MN} B_{\r\s] N} - \frac{3}{16} \e^{MNKLP} B_{[\m\n| N} \dt_{KL} B_{|\r\s] P} \\
& - 32 \e_{N \mP\mQ} \left(A_{[\m}{}^{MN} A_\n{}^\mP \dt_\r A_{\s]}{}^\mQ - \fr14 A_{[\m}{}^{MN} [A_\n, A_\r]_E{}^\mP 
A_{\s]}{}^\mQ\right) \\
& + \frac12 \e^{M K L P Q} \dt_{K L} \mc{G}_{\m\n\r\s P Q},
\end{aligned}
\end{equation}
where $S_{\m\n\r\s P Q}$ is an extra tensor field that drops from the Lagrangian, and the E-bracket is given by
\begin{equation}
    [\L_1, \L_2]_E{}^{MN} = \fr{1}2\left(\mc{L}_{\L_1} \L_2 - \mc{L}_{\L_2} \L_1\right)^{MN}.
\end{equation}
Following \cite{Gubarev:2023kvq}, we consider these field strengths, anholonomy coefficients of the external 7D vielbein on the equal footing with generalized anholonomy coefficients. The latter are usually referred to as generalized fluxes and defined in terms of generalized Lie derivative of the generalized vielbein $E_M{}^{\o{M}}$. Explicitly generalized fluxes read
\begin{equation}
    \begin{aligned}
        \q_{\o{M} \o{N}} & = \fr1{10} \, E \, E^M{}_{[ \o{M}|} \dt_{M N} \left( E^{-1}  E^N{}_{\o{N}]} \right), \\
        Y_{\o{M}  \o{N}} & = - E^M{}_{(\o{M}} \dt_{M N} E^N{}_{\o{N})},\\
        Z_{\o{M}  \o{N} \, \o{K}}{}^{\o{L}} & = E_K{}^{\o{L}}  \partial_{[\o{M}\o{N}} E^K{}_{\o{K}]}  + \frac{1}{3} \left( \, 2 E^M{}_{[\o{M}} \partial_{M N} E^N{}_{\o{N}} + E^M{}_{[\o{M}} E^N{}_{\o{N}} E^{-1} \partial_{M N} E \, \right) \delta_{\o{K}]}{}^{\o{L}}\\
        &  = - \frac{4}{3} \epsilon_{\o{M} \o{N}  \o{K} \o{P} \o{Q}} \, Z^{\o{P} \o{Q},  \o{L}} 
    \end{aligned}
\end{equation}
where we define $\dt_{\o{M}\o{N}} = E^M{_{[\o{M}}} E^N{}_{\o{N}} \dt_{MN}$. Upon Scherk-Schwarz reduction, these reproduce gaugings of the maximal 7D supergravity. Finally, to write the Lagrangian as in \cite{Gubarev:2023kvq} we need fluxes that take into account dependence of the external vielbein on the coordinates on the extended space
\begin{equation} 
\begin{aligned}
    \mc{F}_{M N \mu \nu} & =  e_{\mu}{}^{\o{\rho}} \dt_{M N} e_{\nu \o{\rho}} - \fr{1}7 \, g_{\mu \nu}  e^{-1} \dt_{M N} e \\
    \mc{F}_{M N}{}^{\o{\mu}} &= e^{-\frac{1}{7}} \partial_{M N} \left( e^{\frac{1}{7}} \Omega^{\o{\mu}} \right), \\
    \mc{F}_{\mu \o{M}}{}^{\o{N}} & = E^M{}_{\o{M}} \mc{D}_{\mu} E_{M}{}^{\o{N}} + \fr{1}{14} \delta_{\o{M}}{}^{\o{N}} e^{-1} \mc{D}_{\mu} e ,\\
     \mc{G}_\m & = e^{-1} \mc{D}_{\mu} e  .
\end{aligned} 
\end{equation}
As it has been shown in \cite{Gubarev:2023kvq} the full SL(5) ExFT Lagrangian can be written in terms of the fluxes defined above
\begin{equation}
\begin{aligned}
    e \mc{L}_{Full} & = e \mc{L}_{kin} + e V + \mc{L}_{top} \\
    \mL_{kin} & = -\fr12 \mc{R} -\fr12 \mF^{M N}{}_{(\mu \nu)} \mF_{M N}{}^{(\mu \nu)} -\fr12 \mc{F}^{\mu (M N)} \mc{F}_{\mu (M N)} \\
    & \quad \, - \frac{1}{8} \mc{F}^{\mu \nu}{}_{M N} \mc{F}_{\mu \nu}{}^{M N} + \frac{1}{4\cdot 3!} \mc{F}^{\mu \nu \rho M} \mc{F}_{\mu \nu \rho M} \\
    e^{\frac{2}{7}} \, V &  = - \frac{700}{3}\, \q^{M N} \q_{M N} + Y^{M N} Y_{M N} - \frac{1}{2}\, Y^{M}{}_{M} Y^{N}{}_{N} \\
    & \quad \, + \frac{3}{4}\, 3 Z_{A B C}\,^D \, Z_{G E F}\,^H \, \left( 3 \, \delta^A_H \, \delta^G_D + m_{D H} \, m^{A G} \right) \, m^{B E} \, m^{C F}
\end{aligned}
\end{equation}
where we again skip the topological term $\mc{L}_{top}$ since we are interested in equations of motion.

\subsection{Covariant field equations}
\label{sec:equations}

The framework of polyvector deformations of the scalar truncation of SL(5) ExFT developed in \cite{Gubarev:2020ydf} is based on the following simple idea: to show that a transformation preserves equations of motion it is sufficient to ensure that these equations can be written in terms of fluxes and to show invariance of fluxes. Since all fluxes are expressions of the first order in derivatives of fields, checking their invariance is much easier than doing that for equations of motion, that are second order.  

To derive equations of the SL(5) exceptional field theory in terms of fluxes, let us first list variations of fluxes under general variations of the fundamental fields. For tensor fields it is most convenient to use the so-called covariant variations, that respect tensor hierarchy
\begin{equation}
    \begin{aligned}
        \Delta A_{\mu}{}^{M N} & = \delta A_{\mu}{}^{M N},\\
        \Delta B_{\mu \nu M} &= \delta B_{\mu \nu M} - 2 \epsilon_{M N K L P} A_{[\mu}{}^{N K} \delta A_{\nu]}{}^{L P},\\
     \Delta C_{\mu \nu \rho}{}^{M} & =  \delta C_{\mu \nu \rho}{}^{M} - 3 \delta A_{[\mu}{}^{M N} B_{\nu \rho] N} - 2 \epsilon_{N K L R S} A_{[\mu}{}^{M N} A_{\nu}{}^{K L} \delta A_{\rho]}{}^{R S}.
    \end{aligned}
\end{equation}
\begin{equation} 
    \begin{aligned}
        \delta \mc{F}_{\mu \nu}{}^{M N} &=  2 \mc{D}_{[\mu} \Delta A_{\nu]}{}^{M N} - \frac{1}{16} \epsilon^{M N K L P} \partial_{K L} \Delta B_{\mu \nu P} \\
        \delta \mc{F}_{\mu \nu \rho M} &=  3 \mc{D}_{[\mu} \Delta B_{\nu \rho] M} + 6 \epsilon_{M P Q R S} \mc{F}_{[\mu \nu}{}^{P Q} \Delta A_{\r]}{}^{R S} - \partial_{M N} \Delta C_{\mu \nu \rho}{}^{N} \\
        \delta \mc{F}_{\mu \nu \rho \sigma}{}^{M} &=  4 \mc{D}_{[\mu} \Delta C_{\nu \rho \sigma]}{}^{M} - 6 \mc{F}_{[\mu \nu}{}^{M N} \Delta B_{\rho \sigma] N} + 4 \mc{F}_{[\mu \nu \rho| N} \Delta A_{|\sigma]}{}^{M N}\\
        & + \frac12 \e^{M K L P Q} \dt_{K L} \D \mc{G}_{\m\n\r\s P Q}.
    \end{aligned}
\end{equation}
For scalar fields we define $ \d E^M{}_{\o{M}} E_M{}^{\o{N}} = u^{\o{N}}{}_{\o{M}} =-\d E_M{}^{\o{N}} E^M{}_{\o{M}}$. This implies for variations of fluxes the following
\begin{equation} 
    \begin{aligned}
        \delta \theta_{\o{M} \o{N}} & =   -\frac{1}{10}{\partial}_{\o{K} [\o{M}}{{u}^{\o{K}}\,_{\o{N}]}}\, + \frac{1}{10} {\partial}_{\o{M} \o{N}}{{u}^{\o{K}}\,_{\o{K}}}\,  - 2 {u}^{\o{K}}\,_{[\o{M}} {\q}_{\o{N}] \o{K}},\\
        \delta Y_{\o{M} \o{N}} &=  {\partial}_{\o{K} (\o{M}}{{u}^{\o{K}}\,_{\o{N})}}\,  + 2{u}^{\o{K}}\,_{(\o{M}} Y_{\o{N}) \o{K}},\\
        \d Z^{\o{L} \o{P} \o{Q}} &=  - \frac{1}{16}\, {\partial}_{\o{M} \o{N}}{{u}^{\o{Q}}\,_{\o{K}}}\,  {\epsilon}^{\o{L} \o{P} \o{M} \o{N} \o{K}} - 2{Z}^{\o{L} \o{P} ,[\o{M}} {u}^{\o{Q}]}\,_{\o{M}} - 2{u}^{[\o{P}}\,_{\o{M}}{Z}^{\o{L}] \o{M}, \o{Q}} \\ 
        & \quad \, + \frac{1}{48}\, {\partial}_{\o{M} \o{N}}{{u}^{\o{K}}\,_{\o{K}}}\,  {\epsilon}^{\o{Q} \o{L} \o{P} \o{M} \o{N}} + \frac{1}{24}\, {\partial}_{\o{M} \o{N}}{{u}^{\o{M}}\,_{\o{K}}}\,  {\epsilon}^{\o{Q} \o{L} \o{P} \o{N} \o{K}}. \\
    \end{aligned} 
\end{equation}

Using these, we arrive at the following field equations of the full SL(5) exceptional field theory with $\mF_{\m M N} = \mF_{\m \o{M}}{}^{\o{N}} E_{(M}{}^{\o{M}} E_{N) \o{N}}$, $\mF^{\m \n}{}_{M N} = g^{\m \r} g^{\n \s} \mF_{\r \s}{}^{K L} m_{M K} m_{N L}$ and $\mF^{\m \n \r M} = g^{\m \x} g^{\n \s} g^{\r \psi} \mF_{\x \s \psi N} m^{M N}$
\begin{equation}
\label{eq:equations}
\begin{aligned}
    g_{\mu \nu} & : &         
    \mc{R}_{\mu \nu} - \frac{1}{2} g_{\mu \nu} \mc{R} & = \mc{T}^{(A)}{}_{\mu \nu} + \mc{T}^{(T)}{}_{\mu \nu} + \mc{T}^{(E)}{}_{\mu \nu}\\
    A_{\mu}{}^{M N} & : & 
    e^{-1} \mc{D}_{\mu}{ ( e \mc{F}^{\mu \nu}{}_{M N} ) } & = \mc{J}^{(e)}{}_{M N}{}^{\nu} + \mc{J}^{(T)}{}_{M N}{}^{\nu} + \mc{J}^{(E)}{}_{M N}{}^{\nu} \\
    B_{\mu \nu}{}_{M}&: & e^{-1} \mc{D}_{\rho}{ ( e \mc{F}^{\rho \mu \nu M} ) } & = \mc{J}^{(A,T) \, \mu \nu M} \\
    C_{\mu \nu \rho}{}^{M}&: & 0 & = \partial_{M} \left[ 4! \, e \mc{F}^{\mu \nu \rho M} - \e^{\mu \nu \rho \xi \pi \sigma \tau} \mc{F}_{\xi \pi \sigma \tau}{}^{M} \right] \\
    m^{M N} & : &             
    e^{-1} \mc{D}_{\mu} ( e \mF^{\mu}{}_{M N} ) &= \mc{P}^{(A)}{}_{M N} + \mc{P}^{(T)}{}_{M N} + \mc{P}^{(E)}{}_{M N} 
\end{aligned}
\end{equation}

Note that the equation for $C_{\m\n\r}{}^M$ is solved trivially by the self duality condition for 11D fields. It is worth to mention, that due to generalized Bianchi identities variation w.r.t. $m_{MN}$ is the as the variation w.r.t. $E_M{}^{\o{M}}$, in particular, the latter automatically renders equations symmetric in $\{MN\}$ upon contraction with $E_{\o{M}}{}^Km_{KN}$. An example where this does not hold is provided by the generalized 11D supergravity equations of \cite{Bakhmatov:2022lin}, where the antisymmetric part of the corresponding equations imposes additional constraints on the fields.

The right-hand side of the above equations is written in terms of expressions naturally referred to as energy-momentum tensor and currents. For energy-momentum tensors we have
\begin{equation}
\begin{aligned}
    \mc{T}^{(A)}{}_{\mu \nu} & = \frac{1}{4} \, \left( g^{\rho \sigma} \mc{F}_{\mu \rho}{}^{M N} \mc{F}_{\nu \sigma}{}^{K L} - \frac{1}{4} g_{\mu \nu} \mc{F}^{\rho \s}{}^{M N} \mc{F}_{ \rho \s}{}^{KL} \right) m_{MK} \, m_{NL} \\ 
    \mc{T}^{(T)}{}_{\mu \nu} & = - \frac{1}{2^7} \, \left( \mF_{\mu }{}^{\rho \s}{}_{M} \mF_{\nu \r \sigma N} - \frac{1}{6} g_{\mu \nu} \mF^{\rho  \s \l}{}_{ M} \mF_{\r \s \l N} \right) m^{MN} \\
    \mc{T}^{(E)}{}_{\mu \nu} & = \left( \mc{F}_{\mu (M N)} \mc{F}_{\nu (K L)} - \frac{1}{2} g_{\mu \nu} \mc{F}^{\rho}{}_{(M N)} \mc{F}_{\rho (K L)} \right) m^{MK} \, m^{NL} \\
    & \quad \, + e^{-\frac{8}{7}} \dt_{M N} \left( e^{\frac{8}{7}} m^{MK} \, m^{NL} \mc{F}_{K L \mu \nu} \right) - \frac{5}{7} \, g_{\mu \nu} \, e V
\end{aligned}
\end{equation}
Currents defining the RHS of Maxwell equations for the 2- and 3-form field strengths read
\begin{equation}
\begin{aligned}
    \mc{J}^{(e)}{}_{M N \mu} & = \fr65 \mc{F}_{M N}{}^{\mu} - \fr15\mc{F}_{M N}{}^{{\mu} {\nu}} \Omega_{{\nu}} + \frac{3}{2} \Omega^{[ {\mu} \nu \rho ]} \mc{F}_{M N \n \rho} \\
    \mc{J}^{(T)}{}_{M N}{}^{\mu} & = \fr1{2^7} \, \left( \epsilon_{M N P L K} \mc{F}^{\mu \nu \r P} \mc{F}_{\nu \r}{}^{L K} - 16 \mc{F}^{\mu \nu \rho \s}{}_M \mc{F}_{\nu \rho \s N} \right) \\
    \mc{J}^{(E)}{}_{M N \mu} & = 2 \dt_{K [M}\mc{F}_{\mu N}{}^K  - \frac{5}{2} \dt_{M N} \mc{G}_{\mu} - \frac{50}{9} \left( \mc{F}_{\mu [M}{}^K \q_{N] K} + \mc{G}_{\mu} \q_{M N} \right) \\
    & \quad \, + \frac{1}{2} \mc{F}_{\mu [M}{}^K Y_{N] K} + \mc{F}_{\mu K}{}^L Z_{M N L}{}^K \\
    \mc{J}^{(A,T) \, \mu \nu M} & = -2\epsilon^{M N K L Q} e^{-1} \partial_{N K}{ ( e \mc{F}^{\mu \nu}{}_{L Q} ) } -12 \mc{F}^{\mu \nu \xi \rho}{}_N \mc{F}_{\xi \rho}{}^{N M},
\end{aligned}
\end{equation}
Equations for the generalized metric $m_{MN}$, that is the last line of \eqref{eq:equations}, can also be written in the form of a Maxwell equation for the corresponding 1-form flux. The r.h.s. of these equations are also components of the full 11D energy-momentum tensor, however it is more convenient to refer to them as scalar current. Explicitly they read
\begin{equation}
\begin{aligned}
    \mc{P}^{(A)}{}_{M N} & = \fr12\mc{F}^{\mu \nu}{}_{K M} \mc{F}_{\mu \nu N}{}^{K} \\
    \mc{P}^{(T)}{}_{M N} & = -\fr{1}{3! \cdot 16 } \mc{F}^{\mu \nu \rho}{}_M \mc{F}_{\mu \nu \rho N} \\ 
    \mc{P}^{(E)}{}_{M N} & = \fr12\mc{F}_{K M}{}^{\mu \nu} \mc{F}^{K}{}_{N \mu \nu} + \frac{1}{4} \, E_{M \o{M}} \, E_{N}{}^{\o{N}} \, \mc{P}_{\o{N}}{}^{\o{M}}  \\
    \mc{P}_{\o{N}}{}^{\o{M}} &=  \frac{70}{3} \left( \d^{\o{M}}_{\o{N}} \dt_{\o{K} \o{L}} \q^{\o{K} \o{L}} - \dt_{\o{N} \o{K}} \q^{\o{K} \o{M}} - 20 \, \q_{\o{N} \o{K}} \q^{\o{M} \o{K}} \right) - \partial_{\o{K} \o{L}} Z_{\o{N}}{}^{\o{K} \, \o{L} \, \o{M}} + 12 \, Z_{\o{K} \, \o{L} \, \o{N}} Z^{\o{K} \, \o{L} \, \o{M}} \\
    &\quad \,  - \dt_{\o{N} \o{K}} \left( Y^{\o{M} \o{K}} - \frac{1}{2} \d^{\o{M} \o{K}} Y^{\o{L}}{}_{\o{L}} \right) + 2 \left( Y_{\o{N} \o{K}} Y^{\o{M} \o{K}} - \frac{1}{2} Y^{\o{K}}{}_{\o{K}} Y^{\o{M}}{}_{\o{N}} \right)
    \end{aligned}
\end{equation}

Equations of motion are now written in an SL(5)-covariant form, and the final task is to ensure that each equation transform linearly under a tri-vector generalized Yang-Baxter deformation.  As we discuss in more details in the next section this is an SL(5) transformation, and moreover in \cite{Gubarev:2020ydf} it has already been shown that $\q_{\o{M}\o{N}}$, $Y_{\o{M}\o{N}}$ and $Z^{\o{M}\o{N}, \o{K}}$ (with flat indices!) stay invariant, i.e. indeed transform covariantly. Therefore, to show covariance of all equations we have to i) show that the remaining fluxes transform covariantly, ii) the transformation matrix passes through all derivatives (given certain condition) and eventually multiplies the whole equation.

\section{Three-vector deformations}

Polyvector deformations of 11D backgrounds in a 11=D+d split correspond to generators of the U-duality group $E_{d(d)}$ that have negative weight w.r.t. the $\rmGL(1)$ subgroup of its $\rmGL(d)$ decomposition. For the case in question we have generators $T_{M}{}^{N}$ of the $\rmSL(5)$ group and the following commutation relations
\begin{equation}
    [T_{M}{}^{N}, T_{K}{}^{L}] = \d_M{}^LT_{K}{}^N - \fr15 \d_M{}^N T_K{}^L,
\end{equation}
and the generators are traceless $T_M{}^M = 0$. Under the action of the subgroup $\rmGL(1)\times \rmSL(4) < \rmSL(5)$ the generators decompose according to
\begin{equation}
    \mathbf{24} \to \mathbf{15}_0 + \mathbf{1}_0 +\mathbf{4}_5 + \o{\mathbf{4}}_{-5},
\end{equation}
where the subscript denotes weight w.r.t. the $\rmGL(1)$ subgroup. Generators of non-negative weight upon exponentiation give  the generalized vielbein $E_M{}^{\bar{M}}$, while generators in the $\o{\mathbf{4}}_{-5}$ give the desired tri-vector deformation. In components we have
\begin{equation}
    O_M{}^N = \exp \big[\W^{mnk} T_{mnk}\big]{}_M{}^N,
\end{equation}
where $T_{mnk} = \e_{mnkl}T_5{}^l$. The same construction for larger groups includes more generators, and one has to introduce six-vectors $d=5,6$ and tensors of mixed symmetry for $d=7,8$.

Certainly, although the ExFT action is manifestly invariant under global $\rmSL(5)$ transformations and is invariant under the local generalized Lie derivative, it is in general not invariant under arbitrary local $\rmSL(5)$ transformations. To end up with transformations generalizing bi-vector Yang-Baxter transformations one restricts the tri-vector to have the following form
\begin{equation}
    \W^{mnk} = \fr1{3!} \r^{\a\b\g}k_\a{}^m k_\b{}^n k_\g{}^k.
\end{equation}
Here $k_\a{}^m$ are Killing vectors of the undeformed background and 
\begin{equation}
    [k_\a,k_\b]^m=f_{\a\b}{}^\g k_\g{}^m,
\end{equation}
where the vectors $k_\a{}^m$ generate isometries of the four-dimensional manifold in the usual sense. Let us now show that the same generalized Yang-Baxter equations on constants $\r^{\a\b\g}$ found in \cite{Sakatani:2019zrs,Malek:2019xrf} together with the so-called unimodularity constraint
\begin{equation}
\label{eq:gCYBE}
    \begin{aligned}
        gCYBE&=0,\\
        Uni & = 0
    \end{aligned}
\end{equation}
are sufficient for such a deformation to generate a solution.

\subsection{Sufficient conditions}

Let us consider a general 11D background defined by a metric $G_{\hat{\m}\hat{\n}}$ and a 4-form flux $F_{\hat{\m}\hat{\n}\hat{\r}\hat{\s}}$. Let $\X^{\hat{\m}}$ be an 11-dimensional Killing vector of the background
\begin{equation}
    \label{eq:lie11}
    L^{11}{}_{\Xi} G_{\hat{\mu} \hat{\nu}} = 0 , \quad L^{11}{}_{\Xi} C_{\hat{\mu} \hat{\nu} \hat{\rho}} = 0,
\end{equation}
where $L^{11}$ denotes the full 11D Lie derivative, and we impose the condition on the 3-form potential rather than the flux. This has to be done with certain care due to gauge invariance, however, one can always find such a gauge where the 3-form potential and the 4-form field strength have the same isometries. Under the standard Kaluza-Klein ansatz the Killing vector is decomposed as
\begin{equation}
\label{eq:killing11}
    \Xi^{\hat{\mu}} = ( \xi^{\mu} , k^{m} - \xi^{\nu} A_{\nu}{}^{m} ) .
\end{equation}
Given this and the split form of the 11D fields the equations \eqref{eq:lie11} imply the following
\begin{equation} \begin{aligned} 
     L^{11}{}_{\Xi} G_{\hat{\mu} \hat{\nu}} & = 0  : && \left\{
        \begin{aligned}
          ( \mc{L}^{7}_{\xi} + L^{4}_{k} ) \phi_{m n} & = 0 , \\
          ( \mc{L}^{7}_{\xi} + L^{4}_{k} ) g_{\mu \nu} & = 0 \\
          \mc{D}_{\mu} k^{m} & = ( F_{\mu \nu}{}^m - \phi^{-\fr25} g_{\mu \nu} \partial^{m} ) \xi^{\nu} \\
        \end{aligned}\right. \\
    L^{11}{}_{\Xi} C_{\hat{\mu} \hat{\nu} \hat{\rho}} & = 0 : && \left\{
        \begin{aligned}
          ( \mc{L}^{7}_{\xi} + L^{4}_{k} ) A_{m n k} = & - 3 A_{\mu m n} \partial_{k}{\xi^{\mu}} , \\
          ( \mc{L}^{7}_{\xi} + L^{4}_{k} ) A_{\m m n} = & \left( 2 ( A_{\n \m [m} + 2 A_{[\n}^{l} A_{\m] l [m} ) \delta_{n]}{}^{p} + \phi^{-\fr25} g_{\m \n} A_{m n l} \phi^{l p} \right) \partial_{p}{\xi^{\nu}} ,\\
          ( \mc{L}^{7}_{\xi} + L^{4}_{k} ) A_{\m \n m} = & \left( A_{\r \m \n} + 4 A_\m{}^k A_\n{}^l A_{\r k l} + 6 A_\r{}^p A_\m{}^k A_\n{}^l A_{p k l} \right) \dt_m \x^\r \\ 
          & + \left( 4 A_{[\m}{}^k A_{\n] \r [m} \d_{k]}{}^n - 2 g_{\r [\m} A_{\n] m k} \phi^{-\fr25} \phi^{k n} \right) \dt_n \x^\r \\
          ( \mc{L}^{7}_{\xi} + L^{4}_{k} ) A_{\m \n \r} = & 6 A_{[\m}{}^m ( 2 A_{\n m k} - 3 A_\n{}^n A_{m n k} ) ( \mc{L}^{7}_{\xi} + L^{4}_{k} ) A_{\r]}{}^k \\
          & + 3 \phi^{-\fr25} g_{\s [\m} ( A_{\n \r] m} - 2 A_\n{}^k A_{\r]}{}^l A_{k l m} ) \dt^m \x^\s \\
          & - 12 A_{[\m}{}^m A_\n{}^n ( A_{\r] \s m} - 2 A_{\r]}{}^k A_{\s k m} ) \dt_n \x^\s
        \end{aligned}\right. 
    \label{KilStr}
\end{aligned} \end{equation}   
Here the 7D Lie derivative $\mc{L}^{7}_{\xi}$ is constructed of covariant derivatives $D_\m$ and $L^{4}_k$ is the usual Lie derivative in 4D. For a set of Killing vectors $\X_\a{}^{\hat{\m}}$ commutation relations decompose as
\begin{equation}
    L^{11}{}_{\Xi_{\alpha}} \Xi_{\beta}{}^{\hat{\mu}}  = f_{\alpha \beta}{}^{\gamma} \Xi_{\gamma}{}^{\hat{\mu}} : \quad  \left\{
        \begin{aligned}
            ( \mc{L}^{7}_{\xi_{\alpha}} + L^{4}_{k_{\alpha}} ) \xi_{\beta}{}^{\mu} &= f_{\alpha \beta}{}^{\gamma} \xi_{\gamma}{}^{\mu},\\ L^{4}_{k_{\alpha}} k_{\beta}^{m} & = f_{\alpha \beta}{}^{\gamma} k_{\gamma}{}^{m}
        \end{aligned}\right.
\end{equation}
The last line above suggest that it is natural to define the tri-vector deformation parameter in terms of the vectors $k_\a{}^m$ rather than the combination $k_\a{}^m - \x_\a{}^\m A_\m{}^m$. Explicitly the matrix reads
\begin{equation}
    O^N{}_M = 
        \begin{bmatrix}
            \d_m{}^n & \e_{m n k l} \Omega^{n k l} \\
            0 & 1
        \end{bmatrix}
    \label{DefNot1}
\end{equation}

Next we define transformations of all fields of the SL(5) ExFT that is simply the linear action by the matrix $O_M{}^N$:
\begin{equation} 
\begin{aligned}
    e'{}_{\mu}{}^{\o{\mu}}& = e_{\mu}{}^{\o{\mu}} && & A'{}_{\mu}{}^{M N} & = O^{M}{}_{K} O^{N}{}_{L} A_{\mu}{}^{K L}, \\
    E'{}^{M}{}_{\bar{N}} & = O^{M}{}_{K} E{}^{K}{}_{\bar{N}},  && & B'{}_{\mu \nu M} & = O_{M}{}^{K} B_{\mu \nu K}, \\
    & && & C'{}_{\mu \nu \rho}{}^{M} & = O^{M}{}_{K} C_{\mu \nu \rho}{}^{K}.
    \label{DefNot2}
\end{aligned} 
\end{equation} 
To check that all fluxes are invariant under such a transformation subject to certain algebraic constraints on $\r^{\a\b\g}$ we first note that for the generalized fluxes $\q, Y, Z$ this has already been done in \cite{Gubarev:2020ydf}. Second, we find the non-covariant part of transformations of $\mc{D}_\m$ acting on the external metric and the generalized vielbein
\begin{equation} \begin{aligned}
         \mc{D}'{}_\mu g_{\nu \rho} & = \mc{D}_\mu g_{\nu \rho} - \Delta_{\mu , \nu \rho}, \\
         \mc{D}'{}_\mu E'{}^M{}_A & = O^M{}_K \left( \mc{D}_\mu E^K{}_A - \Delta_\mu{}^K{}_A \right),  \\
        \mc{D}'{}_\mu E'_M{}^A & = O_M{}^K \left( \mc{D}_\mu E_K{}^A - \Delta_{\mu K}{}^A \right).
        \label{DefDg}
\end{aligned} \end{equation}
The non-covariant part of the transformations above read
\begin{equation}
\label{eq:noncov1}
\begin{aligned}
    \Delta_{\mu , \nu \rho} & = \omega_{m n}{}^\gamma \left( A_\mu{}^{m n} L^4_{k_{\gamma}} g_{\nu \rho} + \frac{5}{2} g_{\nu \rho} L^4_{k_{\gamma}} A_\mu{}^{m n} \right) \\
    \Delta_\mu{}^K{}_A & = \begin{bmatrix}
            \omega_{m n}{}^\gamma
        \left( A_\mu{}^{m n} L^4_{k_{\gamma}} E^5{}_A + \frac{2}{5} E^5{}_A L^4_{k_{\gamma}} A_\mu{}^{m n} \right) - E^l{}_A \tilde{R}_{\mu l}\\
            \omega_{m n}{}^\gamma
        \left( A_\mu{}^{m n} L^4_{k_{\gamma}} E^k{}_A + \frac{2}{5} E^k{}_A L^4_{k_{\gamma}} A_\mu{}^{m n} + 2 E^m{}_A L^4_{k_{\gamma}} A_\mu{}^{k n} \right)  
        \end{bmatrix} \\
    \Delta_{\mu K}{}^A & = \begin{bmatrix}
            \omega_{m n}{}^\gamma
        \left( A_\mu{}^{m n} L^4_{k_{\gamma}} E_5{}^A - \frac{2}{5} E_5{}^A L^4_{k_{\gamma}} A_\mu{}^{m n} \right) \\
            \omega_{m n}{}^\gamma
        \left( A_\mu{}^{m n} L^4_{k_{\gamma}} E_k{}^A - \frac{2}{5} E_k{}^A L^4_{k_{\gamma}} A_\mu{}^{m n} + 2 E_l{}^A L^4_{k_{\gamma}} A_\mu{}^{l [m} \delta^{n]}_k \right) + E_5{}^A \tilde{R}_{\mu k} ,
        \end{bmatrix}.
\end{aligned}
\end{equation}
where we have introduced the following notations 
\begin{equation}
\begin{aligned}
    \omega_{m n}{}^{\gamma} & = \frac{1}{4} \epsilon_{m n k l} \, \rho^{\alpha \beta \gamma} k_{\alpha}{}^{k} k_{\beta}{}^{l}, \\
    \tilde{R}_{\mu m} & = A_{\mu}{}^{n k } \tilde{R}_{m n k} - 3 \, \omega_{m n}{}^\gamma D_{\mu}{k_\gamma{}^n}.
\end{aligned}
\end{equation}
The tensor $\tilde{R}_{mnk}\equiv 3 W_{[m}\dt_n W_{k]} $ here is proportional to the R-flux of the SL(5) theory and vanishes upon the generalized Yang-Baxter equation \eqref{eq:gCYBE}. Vanishing of these non-covariant transformations $\D$ guarantees covariant transformation rules for all fluxes, except field strengths for the 1-form and 2-form gauge potentials. For these we have the following transformations rules
\begin{equation} 
\label{eq:transfF}
\begin{aligned}
        & \mc{F} '_{\mu \nu}{}^{M N} = O^M{}_K O^N{}_L \left( \mc{F}_{\mu \nu}{}^{K L} - \Delta_{\mu \nu}{}^{K L} \right), \\
        & \mc{F} '_{\mu \nu \rho M} = O_M{}^K \left( \mc{F}_{\mu \nu \rho K} - \Delta_{\mu \nu \rho K} \right)
\end{aligned} 
\end{equation}
and the non-covariant parts read
\begin{equation}
\label{eq:noncov2}
\begin{aligned}
    \Delta_{\mu \nu}{}^{K L} & = \begin{bmatrix}
            2 \omega_{m n}{}^{\gamma}
        \left( A_\mu{}^{m n} L^4_{k_{\gamma}} A_\nu{}^{k 5} + A_\nu{}^{k 5} L^4_{k_{\gamma}} A_\mu{}^{m n} + 2 A_\nu{}^{m 5} L^4_{k_{\gamma}} A_\mu{}^{n k} \right) - A_\mu{}^{k l} \tilde{R}_{\nu l} \\
            \frac{1}{2} \epsilon^{k l m n} \omega_{m n}{}^\gamma L^4_{k_{\gamma}} B_{\mu \nu 5}
        \end{bmatrix}, \\
    3 \Delta_{\mu \nu \rho K} & = \begin{bmatrix}
            \omega_{m n}{}^\gamma A_\mu{}^{m n} L^4_{k_{\gamma}} B_{\nu \rho 5} \\
            \frac{2}{3} \omega_{k l}{}^\gamma L^4_{k_{\gamma}} C_{\mu \nu \rho}{}^{l} + \omega_{m n}{}^\gamma \left( L^4_{k_{\gamma}} ( A_\mu{}^{m n} B_{\nu \rho k} ) - \frac{8}{3} A_\mu{}^{l 5} \tilde{A}_{\nu l k} L^4_{k_{\gamma}} A_\rho{}^{m n} \right) + B_{\mu \nu 5} \tilde{R}_{\rho k} 
        \end{bmatrix}.
\end{aligned}
\end{equation}
Here we have used the following identity
\begin{equation} 
\begin{aligned}
    [ \mc{D}_{\mu}, L^{4}_{k_{\gamma}} ] = L^{4}_{\mc{D}_{\mu} k_{\gamma} } .
\end{aligned} 
\end{equation}

The expressions for non-covariant terms in the transformations above are completely general and are not subject to any truncation. Let us now derive and analyze the conditions sufficient for all $\D$'s above to vanish. First, recall again the result of \cite{Gubarev:2020ydf} where covariance  of the internal fluxes $\q,Y,Z$ under tri-vector deformations has been shown explicitly given the equations \eqref{eq:gCYBE} hold and $k_\a{}^m$ are Killing vectors for the scalar fields
\begin{equation}
    \begin{aligned}
        L^4_{k_\a} \f_{mn} &=0, && L^4_{k_\a}C_{mnk} = 0.
    \end{aligned}
\end{equation}
This is still true in our case. Next, almost all terms in \eqref{eq:noncov1} and \eqref{eq:noncov2} vanish if we in additional require the same for tensor fields
\begin{equation}
    \begin{aligned}
        L_{k_\a}^4 g_{\m\n} & =0, && L_{k_\a}^4A_\m{}^m =0, \\
        L_{k_\a}^4 A_{\m\n\r} & =0, && L_{k_\a}^4 A_{\m\n m} = 0, && L_{k_\a}^4 A_{\m mn} = 0.
    \end{aligned}
\end{equation}
Finally, to have $\tilde{R}_{\m m}=0$ we need generalized Yang-Baxter equation and
\begin{equation}
\label{eq:covconst}
    D_\m k_\a{}^m = 0.
\end{equation}

Hence, we conclude that the  framework of tri-vector deformations for the truncated ExFT of \cite{Gubarev:2020ydf} extends to the full theory applicable to any background if the deformation is performed along a set of full 11-dimensional Killing vectors $\X_\a{}^{\hat{\m}}$ whose four-dimensional part $k_\a^m$ (as in \eqref{eq:killing11}) is covariantly constant on the external space. Although the general formalism does not require vanishing of the external part $\x^\m$ of the Killing vector, it is convenient to have $\x^\m = 0$ in concrete examples. In this case, the condition \eqref{eq:covconst} simply states that $k^m$ does not depend on $x^\m$, which is most often  the case. 

\subsection{Explicit field transformations}

The equations \eqref{DefNot2} and \eqref{eq:transfF} provide tri-vector transformation rules for the fields of the SL(5) exceptional field theory. However, in this form the transformation rules are not immediately applicable to 11-dimensional backgrounds since one has to first split a solution into external and internal parts, dualise fields into the lowest forms and compose U-duality covariant expressions. As we show below it is possible the perform these manipulations for a general transformation on both the initial and the deformed backgrounds and to rewrite the transformation rules in terms of 11-dimensional component fields, the metric and the 3-form. 

We start with the standard parametrization of the metric and the 4-form field strength
\begin{equation}
    \begin{aligned}
        ds_{11}^2 & = \f^{-\fr25} g_{\m\n}dx^\m dx^\n + \f_{mn} \big(dx^m + A_\m{}^m dx^\m\big)\big(dx^n + A_\n{}^n dx^\n\big), \\
        \hat{F}_{(4)} & = \fr{1}{4!} F_{\hat{\m}\hat{\n}\hat{\r}\hat{\s}} dx^{\hat{\m}}\wedge dx^{\hat{\n}}\wedge dx^{\hat{\r}}\wedge dx^{\hat{\s}} \\
        & = F_{(4)} + F_{(3) m} \wedge dx^m + \fr12 F_{(2) mn}dx^{mn} + \fr1{3!} F_{(1)mnk}dx^{mnk}+ \fr1{4!}F_{mnkl}dx^{mnkl},
    \end{aligned}
\end{equation}
where in the last line we use the notations $dx^{m_1\dots m_p} = dx^{m_1}\wedge \dots \wedge dx^{m_p}$ and $F_{(p)m_1\dots m_k}$ is a $p$-form in the external 7-dimensional space-time with $k$ indices. Decomposition of the corresponding 3-form potential can be written in a similar form
\begin{equation}
    \begin{aligned}
        \hat{C}_{(3)} & = \fr{1}{3!} C_{\hat{\m}\hat{\n}\hat{\r}} dx^{\hat{\m}}\wedge dx^{\hat{\n}}\wedge dx^{\hat{\r}} \\
        & = C_{(3)} + C_{(2) m} \wedge dx^m + \fr12 C_{(1) mn}dx^{mn} + \fr1{3!} C_{mnk}dx^{mnk},
    \end{aligned}
\end{equation}
and in what follows we will be using 
\begin{equation}
    V^m = \fr{1}{3!}\f^{-1}\e^{mnkl}C_{nkl},
\end{equation}
with $\f^2 = \det \f_{mn}$.
It proves convenient to define a tensor
\begin{equation}
    \tW_m = \fr{1}{3!} \f\, \e_{mnkl}\W^{nkl}.    
\end{equation}

Let us now present explicit transformation rules of all the fields above under a tri-vector generalized Yang-Baxter deformation. For the metric degrees of freedom, these read 
\begin{equation}
\label{eq:transf_metric}
    \begin{aligned}
        g'_{\m\n} & = g_{\m\n}, \\
         A_\m'{}^m & = A_\m{}^m - \fr12 \phi^{-1} \epsilon^{m n k l} \tW_{n} A_{\mu k l}, \\
         \phi'_{m n} & = \w^4 \left(\f_{mn} + (1+ V^2) \tW_m \tW_n - 2 \tW_{(m}V_{n)}\right), \\
         \f'{}^{mn} & = \w^{-4}\f^{mn} + \w^2 \left(\tW^2 V^m V^n - \tW^m \tW^n + 2 \tW^{(m} V^{n)}(1 -  V^n \tW_n) \right)\\
         \phi' & = \phi\, \omega^{5},
    \end{aligned}
\end{equation}
where to condense notations we introduce $\omega$ as follows 
\begin{equation}
    \begin{aligned}
        \omega^{-6} & = ( 1 - \tW_{k} V^{k} )^2 + \tW^{2}.
    \end{aligned}
\end{equation}

For the fields descending from the 3-form potential we obtain the following transformation rules
\begin{equation}
    \begin{aligned}
        A_\m{}^m & = A_\m{}^m + \fr12 \W^{mkl} A_{\m kl} \\ 
        A'_{mnk}& = \w^6 \left[A_{mnk}\left(1 - \fr1{3!}A_{pqr} { \W^{pqr}}\right) - {\W_{mnk}} \right]\\
        A'_{\mu m n} & = A_{\mu m n} ,\\
        A'_{\mu \nu m} & = A_{\mu \nu m} - \fr12 \W^{npq} A_{[\m |pq|} A_{\n] mn} + \fr18 W_{m} \tilde{A}_{\mu \nu } , \\
        A'_{\mu \nu \rho} & = A_{\mu \nu \rho} - 3 \, \W^{mkl}A_{[\m |kl|} \left( A_\n{}^n + \fr14 \W^{npq} A_{\n |pq|} \right) A_{\r] mn} + \fr18 W_{m} \tilde{A}_{\mu \nu \rho}{}^m , 
    \end{aligned}
\end{equation}    
and the Lagrange multipliers $\tilde{A}_{\m\n}$ and $\tilde{A}_{\m\n\r}{}^m$ do not transform as expected. Note that the transformations of the gauge potentials include terms with Lagrange multipliers and hence cannot be written exclusively in terms of the initial background fields. To overcome this, we turn to their field strengths and obtain
\begin{equation}
\label{eq:transf_gauge}
    \begin{aligned}
         F'_{\mu \nu}{}^m & = F_{\mu \nu}{}^m - \fr12 \W^{m k l} ( F_{\mu \nu k l} - F_{\mu \nu}{}^n A_{n k l} ) \\
         F'_{m n k l} &= \omega^6 F_{m n k l} \\
        F'_{\mu m n k}& = \omega^3 F_{\mu m n k} + \S_{m n k} \W^{r s t} F_{\mu r s t} \\
         F'_{\mu \nu m n}& = F_{\mu \nu m n} - F_{\mu \nu}{}^{k} A_{m n k} \\
         &\quad + \w^{6} \left[F_{\m\n}{}^k - \fr12 \W^{kpq} \left(F_{\m\n pq} - F_{\m\n}{}^lA_{pql}\right)\right]\cdot \left[A_{mnk}\left(1 - \fr1{3!}A_{pqr}\W^{pqr}\right) - \W_{mnk}\right] ,\\
         F'_{\mu \nu \rho m} & = F_{\mu \nu \rho m} + \fr12 \W^{k l n} ( F_{[\mu \nu |k l|} - F_{[\mu \nu}{}^p A_{p k l} ) A_{\rho] n m} \\
         & - \phi \, W_m \left( \fr{e}{4!} \epsilon_{\mu \nu \rho \sigma \xi \eta \zeta} F^{\sigma \xi \eta \zeta} \phi^{\fr15} + V^n ( F_{\mu \nu \rho n} + F_{[\mu \nu}{}^k A_{\rho] k n} ) \right) \\ 
         F'_{\mu \nu \rho \sigma}& = ( 1 - \phi W_m V^m ) F_{\mu \nu \rho \sigma} - \phi W^m \fr{e}{3!} \epsilon_{\mu \nu \rho \sigma \xi \eta \zeta} ( F^{\xi \eta \zeta}{}_m + F^{[\xi \eta |n|} A^{\zeta]}{}_{n m} ) \phi^{-\fr15}
    \end{aligned}
\end{equation}
Altogether equations \eqref{eq:transf_metric} and \eqref{eq:transf_gauge} provide complete transformation rules for an solution to the 11-dimensional supergravity equations under tri-vector deformations along Killing vectors. Generalized Yang--Baxter equation, unimodularity condition and vanishing of Lie derivative are the sufficient condition for these to generate a solution.

\section{Summary and Discussion}

In this work we extend the formalism of tri-vector deformations constructed in \cite{Gubarev:2020ydf} to general backgrounds of 11D supergravity assuming no truncation and particular form of the metric and the 3-form gauge field. The only restriction is imposed by the formalism of the SL(5) exceptional field theory that is the Killing vectors used must be isometries of a submanifold of dimension at most four. Assuming that all fields of ExFT transform covariantly under a tri-vector deformation, that is an SL(5) transformation, we prove that a solution to 11D supergravity equations is mapped to a solution. The sufficient algebraic condition is precisely the same as in the truncated case, as expected, and is given by the generalized Yang--Baxter equation.

There is a subtle point that comes from the SL(5) covariant structure of the chosen ExFT and hence the necessity to dualize tensor fields to the lowest rank. This results in the presence of Lagrange multipliers (dual gauge fields) in transformations of components of the 3-form gauge field. Explicit check shows that this does not cause a problem when working with gauge invariant objects such as the field strengths. This allows us to derive explicit transformation rules under a tri-vector deformation for 11D supergravity component fields.

Since the developed formalism and derived transformation rules are applicable to any 11D background (modulo the above mentioned restriction on Killing vectors) it allows to consider a more general class of solutions with non-vanishing gauge fields. Of particular interest are deformations of AdS backgrounds along compact isometries of the AdS symmetry group. We believe that one would be able to generate defect AdS solutions to 11D supergravity equations similar to examples found in \cite{Gutperle:2023yrd}. Indeed, say for  $\AdS_7\times \SS^4$  such deformations will break conformal symmetry $SO(2,6)$ to its conformal subgroup $SO(2,2)$, suggesting interpretation in terms of adding defect exactly marginal operators. Such analysis goes beyond the scope of this paper and is reserved for future work.

To go beyond the restriction and to cover Type IIB backgrounds one is interested in extending the formalism to the $E_6$ exceptional field theory. Certainly, no surprises are expected on this way, however Killing vectors will be allowed to be taken along isometries of a 6-dimensional submanifold. 

It would also be interesting to extend the analysis of \cite{Kulyabin:2022yls} to the full SL(5) ExFT considered here and to derive the most general condition for a tri-vector deformation to preserve supersymmetry. 
\section*{Acknowledgments}

This work has been supported by Russian Science Foundation grant RSCF-20-72-10144 and in part by Theoretical Physics and Mathematics Advancement Foundation “BASIS” grant \#23-2-1-5-1.

\appendix

\section{Index conventions}
\label{app:index}

In this paper, we use the following conventions for indices
\begin{equation}
    \begin{aligned}
       &\hat{\mu}, \hat{\nu},\ldots = 1 \dots 11&& \mbox{eleven directions, curved}; \\
       &\hat{\alpha}, \hat{\beta},\ldots = 1 \dots 11&& \mbox{eleven directions, flat}; \\       
       &\mu, \nu, \rho, \ldots = 1,\dots,7 && \mbox{external 7+4 decomposed directions (curved)}; \\
       &\bar{\mu}, \bar{\nu}, \bar{\rho}, \ldots 1,\dots,7 && \mbox{external  7+4 decomposed directions (flat)}; \\  
       &k, l, m, n,  \ldots = 1,\dots,4 && \mbox{internal directions (curved)}; \\       
       &\bar{k}, \bar{l}, \bar{m}, \bar{n}, \ldots = 1,\dots,4 && \mbox{internal directions (flat)}; \\  
       & M, N, K, L, \ldots = 1,\dots, 5 && \mbox{ExFT SL(5) in fundamental representation (curved)};  \\
       & A, B, C, D, \ldots = 1,\dots, 5 && \mbox{ExFT SL(5) in fundamental representation (flat)}.
    \end{aligned}
\end{equation}

\bibliography{bib.bib}
\bibliographystyle{utphys}

\end{document}